\documentclass[aps,prb,twocolumn,groupedaddress,showpacs]{revtex4}
\usepackage{graphicx}

\def\Vec#1{\mbox{\boldmath $#1$}}

\makeatletter
\def\btt#1{\texttt{\@backslashchar#1}}%
\DeclareRobustCommand\bblash{\btt{\@backslashchar}}%
\makeatother

\begin{document}

%\baselineskip=2\normalbaselineskip
%ª'Ps‹ó'«'É'·'éê‡

\title{Field-induced staggered magnetic moment in the quasi-two-dimensional organic Mott insulator $\kappa$-(BEDT-TTF)$_{2}$Cu[N(CN)$_{2}$]Cl} 

\author{F. Kagawa}\altaffiliation[Present Address: ]{Multiferroics  Project (MF), ERATO, Japan Science and Technology Agency (JST), c/o Department of Applied Physics, University of Tokyo, Bunkyo-ku, Tokyo 113-8656, Japan}
\affiliation{Department of Applied Physics, University of Tokyo, Bunkyo-ku, Tokyo 113-8656, Japan}

\author{Y. Kurosaki}\affiliation{Department of Applied Physics, University of Tokyo, Bunkyo-ku, Tokyo 113-8656, Japan}

\author{K. Miyagawa}\affiliation{Department of Applied Physics, University of Tokyo, Bunkyo-ku, Tokyo 113-8656, Japan}

\author{K. Kanoda}\affiliation{Department of Applied Physics, University of Tokyo, Bunkyo-ku, Tokyo 113-8656, Japan}

\date{\today}

\begin{abstract}
	We investigated the magnetism under a magnetic field in the quasi-two-dimensional organic Mott insulator 
$\kappa$-(BEDT-TTF)$_{2}$Cu[N(CN)$_{2}$]Cl through magnetization and $^{13}$C-NMR measurements. 
	We found that in the nominally paramagnetic phase (i.e., above N\'eel temperature) the field-induced local moments 
have a staggered component perpendicular to the applied field.
	As a result, the antiferromagnetic transition well defined at a zero field 
becomes crossover under a finite field. 
	This unconventional behavior is  qualitatively reproduced by the molecular-field calculation for
Hamiltonian including the exchange, Dzyaloshinsky-Moriya (DM), and Zeeman interactions. 
	This calculation also explains other unconventional magnetic features in 
$\kappa$-(BEDT-TTF)$_{2}$Cu[N(CN)$_{2}$]Cl reported in the literature. 
	The present results highlight the importance of the DM interaction in  
field-induced magnetism in a nominally paramagnetic phase, especially in low-dimensional spin systems.	

\end{abstract}


\maketitle

%***********************************************************************
\section{Introduction}

	The quasi-two-dimensional (quasi-2D) organic conductor, $\kappa$-(BEDT-TTF)$_{2}$X, exhibits various phases, depending 
on anion X, temperature and pressure, and have provided a good experimental stage for investigating fundamental 
problems in condensed-matter physics \cite{Ref1} [BEDT-TTF, abbreviated as ET hereafter, is bis(ethylenedithio)tetrathiafulvalene].
	For instance, $\kappa$-(ET)$_{2}$Cu[N(CN)$_{2}$]Br and $\kappa$-(ET)$_{2}$Cu(NCS)$_{2}$ show unconventional
superconductivity at ambient pressure,\cite{Ref2,Ref3} and intensive studies regarding its origin are in 
progress.\cite{Ref4}
	$\kappa$-(ET)$_{2}$Cu[N(CN)$_{2}$]Cl (hereafter abbreviated as $\kappa$-Cl) is a Mott insulator and suitable 
for the study of the bandwidth-controlled Mott transition, because this material undergoes the Mott transition by 
soft pressure ($\sim$ 25 MPa). \cite{Ref4.1} 
	In fact, recent experiments on $\kappa$-Cl using the pressure-sweep technique have revealed fundamental 
aspects of the Mott transition: the first-order transition with a finite-temperature critical endpoint 
($T_{\rm cr}$ $\sim$ 40 K) \cite{Ref5,Ref6,Ref7,Ref8,Ref9} and unconventional critical exponents 
of the Mott criticality.\cite{Ref10} 
	Another interesting aspect in $\kappa$-(ET)$_{2}$X is the magnetism on the frustrated triangular lattice. 
	As will be seen in Sec. II, $\kappa$-(ET)$_{2}$X has an anisotropic triangular lattice, where 
localized moments with antiferromagnetic (AF) correlation are
subject to the so-called spin frustration. 
	In this context, the Mott insulator $\kappa$-(ET)$_{2}$Cu$_{2}$(CN)$_{3}$ is an intriguing material, 
because it has nearly isotropic triangular lattice. 
	In fact, neither the N\'eel order nor spin-gapped behavior are observed down to 32 mK in NMR \cite{Ref11,Ref12} and $\mu$SR measurements.\cite{Ref13} 
	A quantum spin liquid has been suggested as the possible ground state of 
$\kappa$-(ET)$_{2}$Cu$_{2}$(CN)$_{3}$.\cite{Ref14,Ref15}

	In this paper, we focus on the magnetism of the Mott insulator, $\kappa$-Cl. 
	Now, its magnetism is understood as follows: the AF transition occurs at $T_{\rm N}$ $\sim$ 27 K accompanied 
by weak ferromagnetism, which is due to the canting of antiferromagnetically ordered spins.\cite{Ref16}
	This canting is attributed to the Dzyaloshinsky-Moriya (DM) interaction, which is inherent in 
the $\kappa$-(ET)$_{2}$X (see Sec. II). 
	However, further investigations seem to be needed. 
	For instance, Hamad \textit{et al}. found two curious features in $^{1}$H-NMR measurements:\cite{Ref17} 
with increasing magnetic field from 0.58 T to 9.3 T, the peak structure of spin-lattice relaxation rate 
1/$T_{1}$ around $T_{\rm N}$ is strongly suppressed, and the peak temperature (usually regarded as $T_{\rm N}$) increases from $\sim$ 22 K to $\sim$ 28 K, although in conventional antiferromagnets 1/$T_{1}$ is field-insensitive 
and $T_{\rm N}$ is robust (or decreases under strong field). 
	Hamad \textit{et al}.\cite{Ref17} interpreted the field-dependent 1/$T_{1}$ as a consequence of slow spin dynamics due to 
the spin frustration; however, the unexpected increase in the nominal $T_{\rm N}$ with magnetic field 
remains puzzling.

	To get more insight into the magnetism of $\kappa$-Cl, we combined magnetization and NMR measurements. 
	On the basis of the experimental results, we argue that the AF transition at a zero magnetic 
field changes into crossover under a finite field. 
	This statement leads to an intuitively strange consequence that field-induced local spin moments 
in the paramagnetic phase have a staggered component as well as the uniform one, but we actually found the 
non-collinear state under a magnetic field. 
	We also performed the molecular-field analysis of Heisenberg model including the DM and Zeeman interactions 
and succeeded in reproducing qualitatively not only our findings but also the two curious 
features found by Hamad \textit{et al}. \cite{Ref17}
	This means that the above-mentioned puzzling magnetism under magnetic field is the manifestation 
of the interplay between DM and Zeeman interactions. 
	To our knowledge, the importance of DM interaction in a nominally paramagnetic phase has not 
been recognized so far. 
	In this paper, we demonstrate that the DM interaction plays a key role in the magnetism under magnetic 
field even above $T_{\rm N}$, especially in low-dimensional spin systems.

%***********************************************************************
\section{CRYSTAL STRUCTURE OF $\kappa$-(ET)$_{2}$X}

\begin{figure}
\includegraphics[width=8.5cm,height=6.5cm]{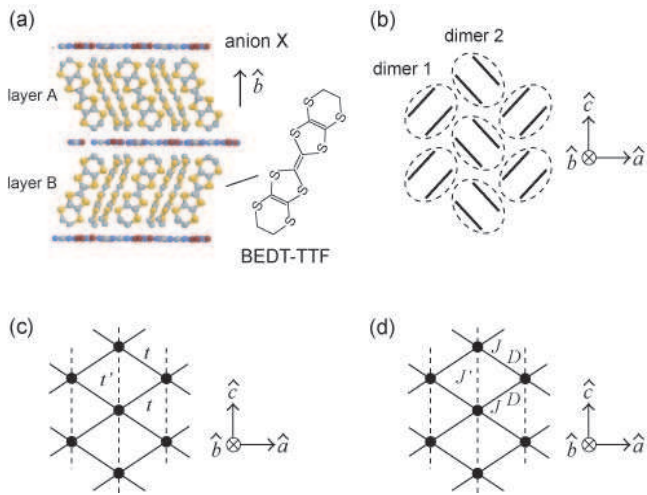}
\caption{\label{Fig1}(Color online) 
	(a) Side view of crystal structure of $\kappa$-Cl. 
	(b) Top view of schematic crystal structure of the conducting ET layer. 
	The pair of lines represents a (ET)$_2$ dimer. 
	(c) Dimer-lattice model of the conducting layer in $\kappa$-Cl. 
	(d) Localized spin model applicable to $\kappa$-Cl.}

\end{figure}

	$\kappa$-(ET)$_{2}$X has a quasi-2D layered structure composed of conducting ET layers 
and insulating anion layers [Fig. \ref{Fig1}(a)]. 
	In the conducting layer, ET molecules form dimers with face-to-face configuration [Fig. \ref{Fig1}(b)]. 
	The monovalent anion X introduces a hole into an antibonding dimer orbital, which forms a two-dimensional 
half-filled band through the inter-dimer transfer integrals. 
	Thus the dimer lattice can be reduced to an isosceles triangular lattice as shown in Fig. \ref{Fig1}(c), 
which is characterized by two inter-dimer transfer integrals, $t$ and $t'$. 
	In the case of the Mott insulating $\kappa$-(ET)$_{2}$X, such as $\kappa$-Cl (space-group $Pnma$), a hole is localized 
at a dimer. 
	Because of the triangular lattice, the localized spins with AF correlation are subject to the spin 
frustration, of which strength is characterized by the ratio $t'/t$. 
	The $t'/t$ of $\kappa$-Cl is estimated at $\sim$ 0.75 (i.e., anisotropic triangular lattice). \cite{Ref11} 
	The unit cell of $\kappa$-Cl extends over two adjacent conducting layers [layer A and layer B, 
see Fig. \ref{Fig1}(a)] and the each layer includes two inequivalent dimers  [dimer 1 and dimer 2, see Fig. \ref{Fig1}(b)]. 
	Thus, the unit cell turns out to contain four inequivalent dimers: A1 and A2 in layer A, and B1 and B2 in layer B.
	These notations are after Ref. 19.
	When the magnetism of $\kappa$-Cl is discussed in terms of the localized spin model, $t$ and $t'$ are 
replaced by AF exchange interactions, $J$ and $J'$, respectively. 
	However, one should keep in mind that the DM interaction also works between the nearest 
dimers [Fig. \ref{Fig1}(d)] because the local inversion symmetry is absent between dimer 1 and dimer 2 [see Fig. \ref{Fig1}(b)].
	In fact, as seen below, the puzzling magnetism found under a magnetic field is understood 
as a consequence of the DM interaction.

%%%%%%%%%%%%%%%%%%%Fig1
%%%%%%%%%%%%%%%%%%%Fig1

%***********************************************************************
\section{EXPERIMENTAL}
\subsection{Magnetization measurements}

	To know the macroscopic magnetism of $\kappa$-Cl, we performed magnetization measurements at 0 
and 7 T with a SQUID magnetometer. 
	In the measurement of spontaneous weak ferromagnetism under 0 T, $\kappa$-Cl was first cooled down to 2 K under 0.1 T (parallel to the conducting layer) and then the magnetic field was switched off.
	Finally, the magnetization along the plane 
was measured in an ascending temperature process under 0 T. 
	The uniform spin susceptibility $\chi_{\rm spin}$ at 7 T was obtained after subtracting the core 
diamagnetism [$-4.7$$\times$$10^{-4}$ emu/(mol f.u.)] from the measured susceptibility, which was determined 
by the magnetization divided by the applied magnetic field.

%%%%%%%%%%%%%%%%%%%%%%%%%%%%%%%%%%%%%%%%%%%%%%%%%%%%%%%%%%%%%%%%%%%%%%%%%%
\subsection{NMR measurements}
\begin{figure}
\includegraphics[width=8.5cm,height=3.8cm]{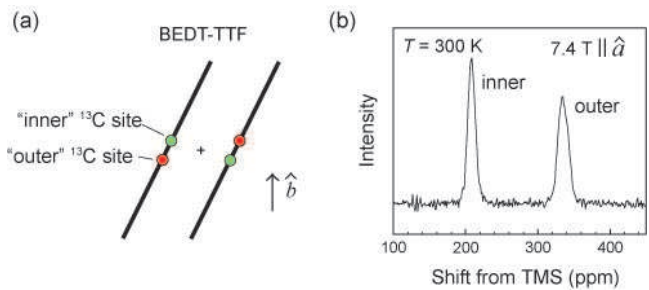}
\caption{\label{Fig2}(Color online) 
	(a) Side view of ET dimer. 
	The cross mark represents the center of the dimer.
	(b) $^{13}$C-NMR spectrum of $\kappa$-Cl under {\boldmath $H$}$\parallel$$\hat{a}$ at room temperature.}

\end{figure}

	In the NMR measurements, we used a $\kappa$-Cl crystal where two central carbon sites in ET molecule 
are substituted by $^{13}$C isotope (hereafter abbreviated as $\kappa$-$^{13}$C$_{2}$-Cl). Under the field applied 
in an arbitrary direction, $\kappa$-$^{13}$C$_{2}$-Cl shows 16 $^{13}$C resonance lines at most. 
	This line profile arises from the following three factors. \cite{Ref19} 
	First, as described above, there are four inequivalent dimers in the unit cell: A1, A2, B1, and B2. 
	Second, the inclined face-to-face geometry of ET molecules in the dimer makes the two central $^{13}$C sites 
in ET inequivalent [Fig. \ref{Fig2}(a)]: the $^{13}$C site closer to the center of the dimer is labeled with ``inner", 
and the other $^{13}$C is labeled with ``outer". 
	Finally, a nuclear dipolar splitting between the central $^{13}$C sites doubles the number of lines. 
	Below we deal mainly with the outer site, since the NMR properties of the inner and outer sites 
show the similar behavior.

	In the present NMR measurements, the magnetic field {\boldmath $H$} of 7.4 T was applied 
parallel to the $\hat{a}$ axis. 
	In that case, the NMR shift of the outer site at dimer $i$ (= A1, A2, B1, B2), $K^{a}_{i}$ (in ppm), is given by 
\begin{eqnarray}
	K^{a}_{i} &=& A^{aa}_{i}\frac{M^{a}_{i}}{H} + A^{ab}_{i}\frac{M^{b}_{i}}{H} + A^{ac}_{i}\frac{M^{c}_{i}}{H} + C^{a}_{i},
\end{eqnarray}
	where $M^{\alpha}_{i}$ denotes the $\alpha$-axis ($\alpha$ $= a, b, c$) component of the local spin moment {\boldmath $M$}$_{i}$ at 
dimer $i$, $A^{a\alpha}_{i}$ is a component of the hyperfine coupling tensor of the outer site in dimer $i$, 
and $C^{a}_{i}$ (in ppm) is the chemical shift of the site. 
	As discussed by Smith \textit{et al},\cite{Ref18} $A^{a\alpha}_{A1}$ is positive for every $\alpha$, 
and the hyperfine coupling constants at other dimers can be determined by applying the symmetry operation 
inherent in $\kappa$-Cl to $A^{a\alpha}_{A1}$ (space-group $Pnma$). 
	The derived relations between the hyperfine coupling constants at different dimers are as follows:
\begin{eqnarray}
	A^{aa}_{A1} =  A^{aa}_{A2} = A^{aa}_{B1} = A^{aa}_{B2} > 0,\\
	A^{ab}_{A1} =  A^{ab}_{A2} = - A^{ab}_{B1} = - A^{ab}_{B2} > 0,\\
	A^{ac}_{A1} =  - A^{ac}_{A2} = A^{ac}_{B1} = -A^{ac}_{B2} > 0. 
\end{eqnarray}

%%%%%%%%%%%%%%%%%%%Fig2
%%%%%%%%%%%%%%%%%%%Fig2

	Under {\boldmath $H$}$\parallel$$\hat{a}$, the dipolar splitting nearly vanishes because {\boldmath $H$} 
approximately forms the so-called magic angle ($\sim$ 54.7$^\circ$) against the $^{13}$C=$^{13}$C vector in ET by happenstance. 
	At room temperature (in the paramagnetic state), all dimers are equivalent under {\boldmath $H$}$\parallel$$\hat{a}$; 
thus $^{13}$C-NMR spectra of $\kappa$-$^{13}$C$_{2}$-Cl shows only two lines coming from the inner $^{13}$C and 
outer $^{13}$C [Fig. \ref{Fig2}(b)]. 
	A marked feature in the experimental results is that the two-line shape holds even at low temperatures 
where the spins are ordered. 
	This means that $K^{a}_{A1} =  K^{a}_{A2} = K^{a}_{B1} = K^{a}_{B2}$ is satisfied in the whole temperature range. 
	By using Eqs. (1)-(4), this relation is rewritten as follows: 
\begin{eqnarray}
	M^{a}_{A1} =  M^{a}_{A2} = M^{a}_{B1} = M^{a}_{B2},\\
	M^{b}_{A1} =  M^{b}_{A2} = - M^{b}_{B1} = - M^{b}_{B2},\\
	M^{c}_{A1} =  - M^{c}_{A2} = M^{c}_{B1} = -M^{c}_{B2}. 
\end{eqnarray}
	Equations (5)-(7) put strict constraints on the spin configuration 
under {\boldmath $H$}$\parallel$$\hat{a}$. 
	In Sec. IV, we determine the spin structure at low temperatures on the basis of these constraints.
	The absolute values of the Knight shift in the present study are calibrated with the room-temperature 
line position of the outer site, which is determined to be 336.2 ppm by Smith \textit{et al}. \cite{Ref18}

%***********************************************************************
\section{EXPERIMENTAL RESULTS}

%%%%%%%%%%%%%%%%%%%Fig3
\begin{figure}
\includegraphics[width=8.5cm,height=6.7cm]{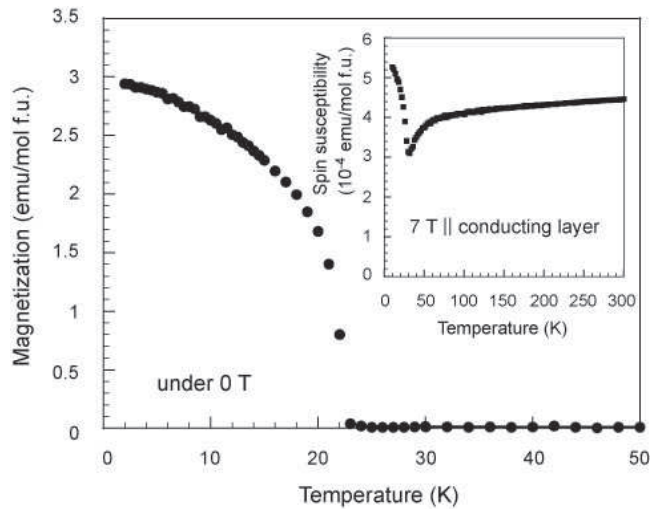}
\caption{\label{Fig3} 
	Temperature dependence of magnetization under a zero magnetic field.
Inset: Temperature dependence of spin susceptibility measured under 7 T (parallel to the conducting layer).}

\end{figure}
%%%%%%%%%%%%%%%%%%%Fig3

	In this section, by investigating the macroscopic and microscopic magnetisms of $\kappa$-Cl, we demonstrate 
(i) the increasing ``$T_{\rm N}$" with a magnetic-field and (ii) the field-induced non-collinear local moment 
in a paramagnetic phase.
	First we show the macroscopic magnetism. 
	The temperature dependence of $\chi_{\rm spin}$ of a $\kappa$-Cl single crystal measured at 7 T 
(parallel to the conducting layer) is presented in the inset of Fig. \ref{Fig3}. 
	The result is consistent with $\chi_{\rm spin}$ of the powdered sample measured at 0.1 T.\cite{Ref20} 
	The weak ferromagnetism observed below 30 K is due to the DM interaction and therefore should appear spontaneously even under a zero field. 
	In fact, as shown in the main panel of Fig. \ref{Fig3}, the magnetization measurements at a zero field confirm a spontaneous 
weak ferromagnetism parallel to the conducting plane below $\sim$ 23 K. 
	This temperature is regarded as $T_{\rm N}$ at a zero field (abbreviated as the zero-field 
$T_{\rm N}$ hereafter) and is in good agreement with the peak temperature of 1/$T_{1}$ under 0.58 T.\cite{Ref17}
	Note that the zero-field $T_{\rm N}$ is appreciably lower than the peak temperature under 3.7 T 
($\sim$ 27 K).\cite{Ref16} 
	Thus, the present results support the finding by Hamad \textit{et al}.\cite{Ref17} that ``$T_{\rm N}$" seems to 
increase with a magnetic field  (actually, as seen below, the peak temperature of 1/$T_{1}$ 
should be regarded as crossover temperature rather than ordering temperature in $\kappa$-Cl).

\begin{figure}
\includegraphics[width=8.6cm,height=8.0cm]{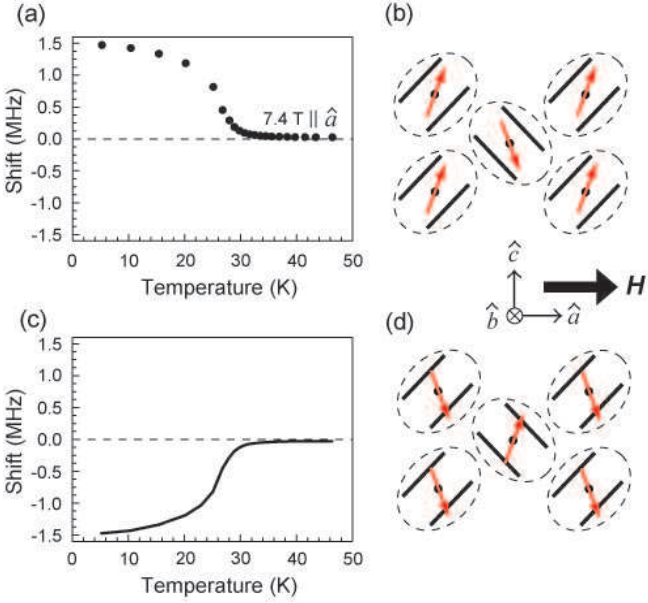}
\caption{\label{Fig4}(Color online) 
	(a) Experimental $^{13}$C-NMR shift of the outer site under 7.4 T parallel to the $\hat{a}$ axis. 
	(b) Spin structure at low temperatures deduced from (a). 
	(c) Hypothetical $^{13}$C-NMR shift of the outer site and (d) spin structure at low temperatures deduced from (c). 
	The horizontal lines in (a) and (c) represent the shift origin. 
	In (b) and (d), the canting of local moments toward the $\hat{a}$ axis is exaggerated.}

\end{figure}

	Next, we turn to the microscopic magnetism of $\kappa$-Cl. 
	Figure \ref{Fig4}(a) shows the temperature dependence of $^{13}$C-NMR shift at the outer site. 
	The large shift toward low temperatures reflects the growth of local moment, consistent with the previous 
report.\cite{Ref18} 
	As demonstrated by Smith \textit{et al},\cite{Ref18} the spin configuration at low temperatures can be 
determined by considering this shift in the light of Eqs. (1)-(7). 
	We describe this procedure in more detail, which leads us to conclude the 
field-induced non-collinear state, one of the findings in the present study. 
	Since the applied field of 7.4 T ($\parallel$$\hat{a}$) is far greater than the spin-flop field,\cite{Ref16} the local 
moments lie nearly in the $bc$ plane with a slight canting toward the field direction 
($\parallel$$\hat{a}$); moreover, the AF interaction between A1 and A2 makes the $b$ and $c$ components of 
local moments antiparallel; thus $M^{b}_{A1} =  - M^{b}_{A2}$ and $M^{c}_{A1} = - M^{c}_{A2}$. 
	From these constraints and Eq. (6), it turns out that the local moments cannot have the $b$ component, 
i.e., $M^{b}_{A1} =  M^{b}_{A2} = 0$. 
	Thus $K^{a}_{A1}$ [see Eq. (1)] is given by
\begin{eqnarray}
	K^{a}_{A1} &=& A^{aa}_{A1}\frac{M^{a}_{A1}}{H} + A^{ac}_{A1}\frac{M^{c}_{A1}}{H} + C^{a}_{A1}.
\end{eqnarray}
	While the sign of magnetization component parallel to the field, $M^{a}_{A1}$, is obviously positive, 
that of the perpendicular component, $M^{c}_{A1}$, is not trivial.
	However, it can be judged explicitly from the sign of the shift at low temperatures, where $|M^{c}_{A1}|$ $\gg$ $M^{a}_{A1}$ $>$ 0 and therefore the second term is dominant in Eq. (8).
	Because $A^{ac}_{A1}$ $>$ 0 as mentioned above (Sec. I\hspace{-.1em}I\hspace{-.1em}I B), 
the large positive shift below $T_{\rm N}$ is attributed to the growth of positive $M^{c}_{A1}$. 
	In this way, all components of the local moment at dimer A1 at low temperatures are determined to be 
$M^{c}_{A1}$ $\gg$ $M^{a}_{A1}$ $>$ 0 and $M^{b}_{A1}$ = 0. 
	By applying Eqs. (5)-(7) to {\boldmath $M$}$_{A1}$, one finally obtains the whole spin configuration
at low temperatures, which is roughly sketched in Fig. \ref{Fig4}(b).

%%%%%%%%%%%%%%%%%%%Fig4
%%%%%%%%%%%%%%%%%%%Fig4

	Figure \ref{Fig4}(a) represents an anomalous feature overlooked in the previous studies: 
on cooling, the NMR spectra always show the positive shift, i.e., the negative shift as shown in 
Fig. \ref{Fig4}(c) has never been observed. 
	This indicates that at low temperatures the spin configuration having the negative $M^{c}_{A1}$ 
as shown in Fig. \ref{Fig4}(d) never appears. 
	Note that such situation is not expected in the conventional AF transition under a magnetic field, 
where the Figs. \ref{Fig4}(b) and \ref{Fig4}(d) configurations are degenerate and thus have the equal chance 
to emerge below $T_{\rm N}$. 
	Therefore, the present result implies that the former configuration has a lower energy than the latter. 
	As seen in Sec. V, the origin of this energy difference is actually understood by considering the DM 
interaction with the antisymmetric nature, i.e. {\boldmath $D$}$_{12}$ $\cdot$ ({\boldmath $S$}$_{1}$$\times${\boldmath $S$}$_{2}$) 
$\neq$ {\boldmath $D$}$_{12}$ $\cdot$ ({\boldmath $S$}$_{2}$$\times${\boldmath $S$}$_{1}$). 
	The point is that the thermal average of the two configurations yields positive $M^{c}_{A1}$ (staggered component) as well as positive $M^{a}_{A1}$ {\it regardless of temperature} because of the energy difference, whatever the origin is. 
	In this way, the inevitable positive shift in Fig. \ref{Fig4}(a) turns out to indicate that the field-induced 
local moments  under {\boldmath $H$}$\parallel$$\hat{a}$ are not uniform 
($M^{a}_{i}$ $>$ 0 and $M^{c}_{i}$ = 0) but non-collinear 
($M^{a}_{i}$ $>$ 0 and $M^{c}_{A1}$ = $- M^{c}_{A2}$ $>$ 0) even above zero-field $T_{\rm N}$.

%%%%%%%%%%%%%%%%%%%Fig5
\begin{figure}
\includegraphics[width=8.4cm,height=8.9cm]{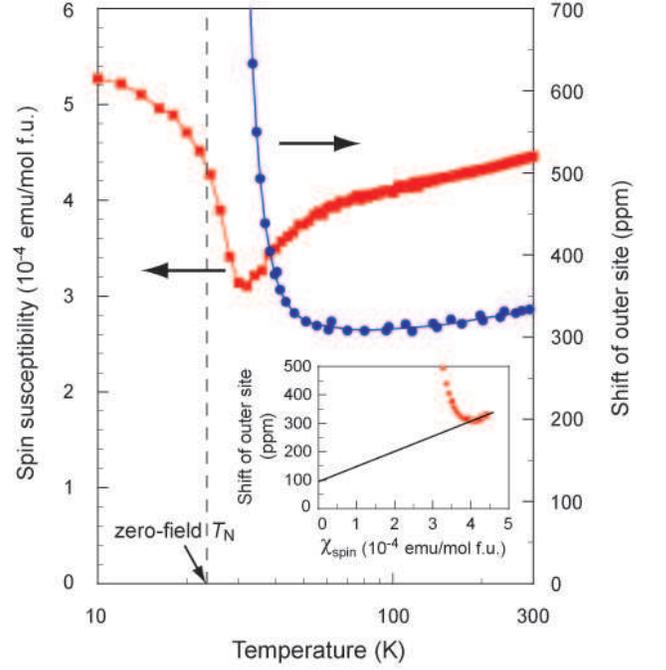}
\caption{\label{Fig5}(Color online) 
	Spin susceptibility measured under 7 T and $^{13}$C-NMR shift of the outer site.
	Solid lines are guide for the eyes.
	Inset: $K$-$\chi$ plot. 
	The hyperfine coupling constant was estimated from the linear fitting shown in the inset.}

\end{figure}
%%%%%%%%%%%%%%%%%%%Fig5

%
	The non-collinear spin state can be identified more evidently when the uniform $\chi_{\rm spin}$ is compared with 
the NMR shift $K^{a}_{A1}$. 
	Figure \ref{Fig5} shows the temperature dependence of NMR shift ({\boldmath $H$}$\parallel$$\hat{a}$) and 
in-plane $\chi_{\rm spin}$. 
	We assume that the in-plane anisotropy of $\chi_{\rm spin}$ is negligibly small, as is the case in most of the 
organic conductors. 
	Therefore, the uniform $\chi_{\rm spin}$ can be interpreted as probing $M^{a}_{A1}$. 
	Note that when the field-induced local moments under {\boldmath $H$}$\parallel$$\hat{a}$ are uniform 
($M^{a}_{i}$ $>$ 0 and $M^{c}_{i}$ = 0) as usual, $\chi_{\rm spin}$ and $K^{a}_{A1}$ should show the same 
temperature dependence, because $K^{a}_{A1}$ is given by $A^{aa}_{A1}M^{a}_{A1}/H$ + $C^{a}_{A1}$ in this case
 [see Eq. (8)].
	As seen in Fig. \ref{Fig5}, however, the two quantities exhibit quite different temperature dependence above the 
zero-field $T_{\rm N}$. 
	Especially, the difference is pronounced in 30-60 K: the shift increases on cooling, indicating the 
growth of local magnetization, while the uniform $\chi_{\rm spin}$ (i.e., $M^{a}_{A1}$) decreases. 
	This remarkable disagreement demonstrates that the finite staggered moment $M^{c}_{A1}$ (= $-M^{c}_{A2}$ $>$ 0) contributes to
the observed shift in 30-60 K.
%

%%%%%%%%%%%%%%%%%%%Fig6
\begin{figure}
\includegraphics[width=7.4cm,height=12.8cm]{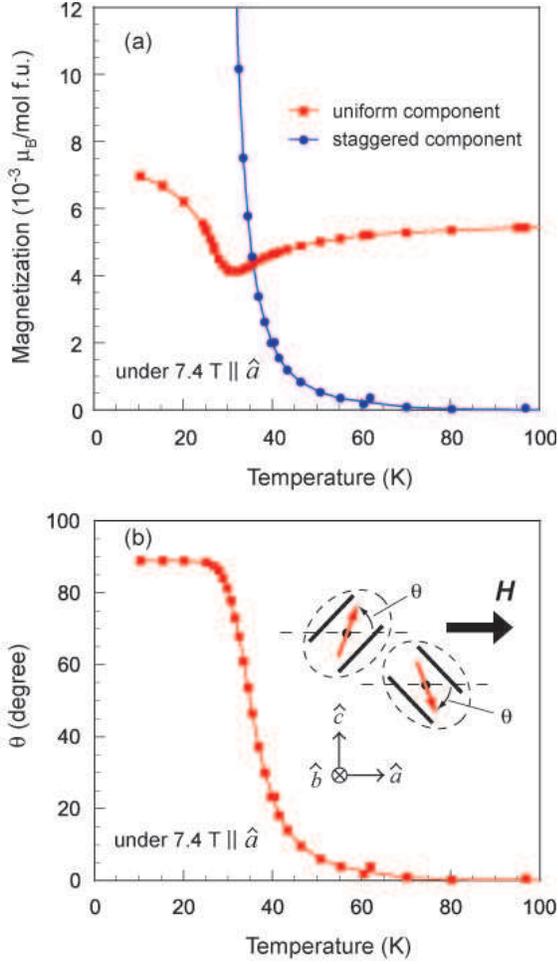}
\caption{\label{Fig6}(Color online) 
	Estimated temperature dependence of (a) the uniform and staggered moment under 7.4 T and (b) the 
non-collinear angle $\theta$ measured from the magnetic-field axis.
	Eye guides are shown together.}

\end{figure}
%%%%%%%%%%%%%%%%%%%Fig6

	Using the reported data of $A^{ac}_{A1} = 1.26A^{aa}_{A1}$ and $C^{a}_{A1} =$ 96 (ppm),\cite{Ref18,Ref19} we can estimate 
the magnitude of staggered moment semi-quantitatively as follows. 
	Here, $K^{a}_{A1}$ [ppm] is given by $A^{aa}_{A1}(M^{a}_{A1}/H + 1.26M^{c}_{A1}/H) +$ 96.
	At high temperatures, it is plausible that the field-induced local moment is nearly parallel to the 
magnetic-field (i.e., $M^{a}_{A1}$ $\gg$ $M^{c}_{A1}$).
	Therefore $K^{a}_{A1}$ around room temperature is given by $A^{aa}_{A1}M^{a}_{A1}/H +$ 96, and 
$A^{aa}_{A1}$ can be estimated roughly from the $K$-$\chi$ 
plot of the high-temperature data, as shown in the inset of Fig. \ref{Fig5}.
	Note again that the difference between the observed shift and $A^{aa}_{A1}M^{a}_{A1}/H +$ 96 
(i.e., the straight line in the inset) is attributed to the contribution of staggered component, 
$A^{ac}_{A1}M^{c}_{A1}$ ($=$ 1.26$A^{aa}_{A1}M^{c}_{A1}$); thus, using the obtained $A^{aa}_{A1}$, one can evaluate
the temperature dependence of the staggered component $M^{c}_{A1}$ as well as that of the uniform component $M^{a}_{A1}$. 
	The results are shown in Fig. \ref{Fig6}(a), where the staggered component $M^{c}_{A1}$ (= $-M^{c}_{A2}$)
starts to grow markedly below $\sim$ 60 K. 
	The temperature dependence of non-collinear angle $\theta$ defined by $\arctan$($M^{c}_{A1}/M^{a}_{A1}$) is  
shown in Fig. \ref{Fig6}(b), which demonstrates that with lowering temperature the field-induced non-collinear local moments gradually tilt in accordance with the growth of staggered component.

	Note again that the positive staggered moment is induced by a magnetic field even above ``$T_{\rm N}$''. 
	Therefore the AF symmetry breaking occurs in the whole temperature range under a field, although its degree
is expected to vanish asymptotically toward high temperatures. 
	Thus, the rapid increase in the staggered moment, which has so far been addressed nominally as 
the AF transition, is not a phase transition but a crossover without thermodynamic singularity. 
	As seen in Sec. V, the crossover behavior under a magnetic field 
is also demonstrated in the molecular-field calculations.

%%%%%%%%%%%%%%%%%%%Fig7
\begin{figure}
\includegraphics[width=8.5cm,height=3.8cm]{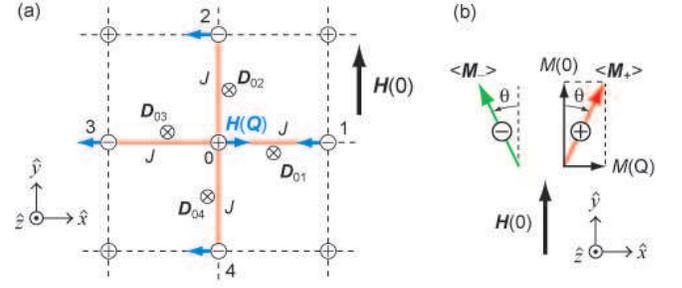}
\caption{\label{Fig7}(Color online) 
	(a) Schematic illustration of the present square-lattice model in the presence of uniform and staggered fields. 
	The local moment at site 0 interacts with four nearest-neighbor sites, 1-4. 
	$+$ and $-$ denote the two magnetic sublattices (see the text).
	(b) Definition of the uniform moment $M$(0), staggered moment $M$({\boldmath $Q$}) and non-collinear angle $\theta$.}

\end{figure}
%%%%%%%%%%%%%%%%%%%Fig7

%%%%%%%%%%%%%%%%%%%Fig8
\begin{figure*}
\includegraphics[width=17cm,height=5cm]{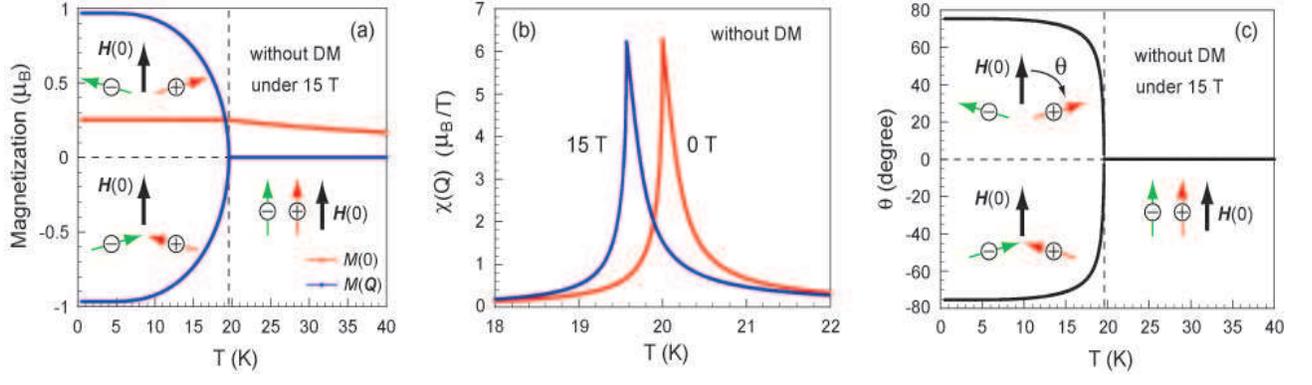}
\caption{\label{Fig8}(Color online) 
	Molecular-field calculation of temperature dependence of (a) the uniform magnetization $M$(0) and the 
staggered magnetization $M$({\boldmath $Q$}) under 15 T, (b) the staggered susceptibility $\chi$({\boldmath $Q$}) 
under 0T and 15 T, and (c) the non-collinear angle $\theta$ under 15T. 
	In the present calculations, the DM interaction is not included.}

\end{figure*}
%%%%%%%%%%%%%%%%%%%Fig8

%*****************************************************************************************************
\section{NUMERICAL RESULTS
---MOLECULAR-FIELD APPROXIMATION}
\subsection{Model Hamiltonian}

	To reveal the origin of the unconventional magnetism highlighted above, we 
investigate the S = 1/2 AF Heisenberg model including the DM and Zeeman interactions on a two-dimensional 
square lattice [Fig. \ref{Fig7}(a)]. 
Here,
\begin{eqnarray}
	 \mathcal{H} =  J\sum_{<i.j>}^{} \Vec{S}_{i} \cdot \Vec{S}_{j} + g\mu_B\sum_{i}^{} \Vec{S}_{i} \cdot 
\Vec{H}_{i} \nonumber \\
+ \sum_{<i.j>}^{} \Vec{D}_{ij} \cdot (\Vec{S}_{i} \times \Vec{S}_{j}), 
\end{eqnarray}
	where $\Vec{S}_{i}$ and {\boldmath $H$}$_{i}$ are the spin operator and external magnetic 
field, respectively, at site $i$, and $<...>$ denotes the sum over the nearest-neighbors. 
	For simplicity, the DM vector {\boldmath $D$} is set to be parallel to the $\hat{z}$ axis, and we 
ignore the AF exchange between the next-nearest-neighbors. 
	A uniform field {\boldmath $H$}({\boldmath $q$} = 0) is applied 
along the $\hat{y}$ axis, where {\boldmath $q$} is a wave-number vector. 
	In the case that only the AF exchange term is present (i.e., {\boldmath $H$}$_{i}$ = 
{\boldmath $D$}$_{ij}$ = 0), the N\'eel order with {\boldmath $Q$} = ($\pi$, $\pi$) is emergent at low 
temperatures; thus the square lattice can be divided into two magnetic sublattices, $+$ and $-$, 
as shown in Fig. \ref{Fig7}(a). 
	We choose the $\hat{x}$ axis as the easy axis, but the magnetic anisotropy is not 
taken into account explicitly.

	In this section, we solve the present model using the molecular-field approximation on the two 
magnetic sublattices. 
	The molecular-field Hamiltonian of Eq. (9) is given by
\begin{eqnarray}
	 \mathcal{H} =  g\mu_B\sum_{i+}^{} \Vec{S}_{i+} \cdot \Vec{H}_+^{\rm eff} + g\mu_B\sum_{i-}^{} \Vec{S}_{i-} \cdot \Vec{H}_-^{\rm eff}.  
\end{eqnarray}
	Here $i+$ ($i-$) denotes a lattice point on the $+$ ($-$) magnetic sublattice, and {\boldmath $H$}$_+^{\rm eff}$ 
({\boldmath $H$}$_-^{\rm eff}$) is the effective field perceived by the $+$ ($-$) magnetic sublattice. 
	The effective field is described by the sum of the external and molecular fields:
\begin{eqnarray}
	 \Vec{H}_+^{\rm eff} =  - \frac{ZJ}{(g\mu_B)^2}\langle \Vec{M}_- \rangle - \frac{Z}{(g\mu_B)^2}\langle 
\Vec{M}_- \rangle \times \Vec{D}_{+-} + \Vec{H}_i, \nonumber \\ \\
	\Vec{H}_-^{\rm eff} =  - \frac{ZJ}{(g\mu_B)^2}\langle \Vec{M}_+ \rangle - \frac{Z}{(g\mu_B)^2}\Vec{D}_{+-} \times \langle \Vec{M}_+ \rangle  + \Vec{H}_i. \nonumber \\
\end{eqnarray}
	Here {\boldmath $D$}$_{+-}$ is along the $\hat{z}$ axis, $Z$ = 4 is the number of the nearest-neighbors, 
and $\langle${\boldmath $M$}$_+ \rangle$ ($\langle${\boldmath $M$}$_- \rangle$) denotes the statistically averaged local moments on the + ($-$) sublattices. 
	In the molecular-field solutions, $\langle${\boldmath $M$}$_+ \rangle$ and $\langle${\boldmath $M$}$_- \rangle$ are determined self-consistently so as to satisfy $\langle${\boldmath $M$}$_+ \rangle$$\parallel${\boldmath $H$}$_+^{\rm eff}$ and $\langle${\boldmath $M$}$_- \rangle$$\parallel${\boldmath $H$}$_-^{\rm eff}$.
	Note that because of Zeeman and DM terms such local moments are symmetric with respect to the 
$\hat{y}$ axis [$\parallel${\boldmath $H$}(0)] and lie in the $\hat{x}$-$\hat{y}$ plane. 
	For clarity, we adopt the basis composed of the uniform moment along the $\hat{y}$-axis, $M$(0) = $|\langle${\boldmath $M$}$_+ \rangle$ + $\langle${\boldmath $M$}$_- \rangle|$/2, and the staggered moment along the $\hat{x}$-axis, $M$({\boldmath $Q$}) = $|\langle${\boldmath $M$}$_+ \rangle$ $-$ $\langle${\boldmath $M$}$_- \rangle|$/2, 
as illustrated in Fig. \ref{Fig7}(b). 
	We define the non-collinear angle $\theta$ as $\arctan$$^{-1}$[$M$({\boldmath $Q$})/$M$(0)].
	When we calculate the staggered susceptibility $\chi$({\boldmath $Q$}) under {\boldmath $H$}(0), 
a staggered field {\boldmath $H$}({\boldmath $Q$}) is applied additionally along the $\hat{x}$ axis [see Fig. \ref{Fig7}(a)]; 
then $\chi$({\boldmath $Q$}) is obtained from the difference of $M$({\boldmath $Q$}) between under {\boldmath $H$}(0) 
and under {\boldmath $H$}(0)$+${\boldmath $H$}({\boldmath $Q$}), as represented in the following equation:
\begin{eqnarray}
	 \chi(\Vec{Q}) \equiv  \frac{M(\Vec{Q}, \Vec{H}(0)+\Vec{H}(\Vec{Q})) - M(\Vec{Q}, \Vec{H}(0))}
{|\Vec{H}(\Vec{Q})|}. 
\end{eqnarray}
	Although $\chi$({\boldmath $Q$}) should be calculated under an infinitesimal 
{\boldmath $H$}({\boldmath $Q$}), the present calculation of $\chi$({\boldmath $Q$}) was performed under 
$|${\boldmath $H$}({\boldmath $Q$})$|$ = 20 mT because of the finite numerical accuracy.
	Below, within the molecular field approximation, we show the temperature dependence of $M$(0), 
$M$({\boldmath $Q$}) and $\chi$({\boldmath $Q$}) under various magnitudes of {\boldmath $H$}(0) and 
{\boldmath $D$}$_{+-}$. 
	Throughout the calculation, $J$ = 20 K and $g$ = 2 were used. 
	The molecular-field $T_{\rm N}$ in the square lattice under a zero field is given by 
$\sqrt{J^{2} + |\Vec{D}_{+-}|^{2}}$. 
	Hence in the case of $|${\boldmath $D$}$_{+-}$$|$ = 0, $T_{\rm N}$ equals $J$, namely, 20 K.

%%%%%%%%%%%%%%%%%%%%%%%%%%%%%%%%%%%%%%%%%%%%%%%%%%%%%%%%%%%%%%%%%%%%%%%%%%
\subsection{Heisenberg model without DM interaction under magnetic field}

	It is instructive to show first the consequences of the Heisenberg model without the DM interaction under a magnetic field. 
	As is well known, the AF transition, which is characterized by the emergence of $M$({\boldmath $Q$}) and the
divergence of $\chi$({\boldmath $Q$}), is still well-defined even under a uniform field of moderate magnitude. 
	This feature is reproduced in our numerical calculation: under 15 T, finite $M$({\boldmath $Q$}) appears 
below $T_{\rm N}$ ($\sim$ 19.57 K) [Fig. \ref{Fig8}(a)], and $\chi$({\boldmath $Q$}) is divergent at 
this temperature [Fig. \ref{Fig8}(b)], although $T_{\rm N}$ is suppressed slightly from 20 K.\cite{Ref21} 
	Below $T_{\rm N}$, $M$({\boldmath $Q$}) can either be positive or negative (i.e., parallel or antiparallel to 
the $\hat{x}$ axis), because the positive- and the negative-$M$({\boldmath $Q$}) 
phases are degenerate. 
	The presence of this spontaneous symmetry breaking also assures the well-defined AF transition 
even under 15 T. 
	Reflecting the well-defined AF transition, the non-collinear state (i.e., finite $\theta$) is emergent only below $T_{\rm N}$, as shown in Fig. \ref{Fig8}(c).
%

%%%%%%%%%%%%%%%%%%%Fig9
\begin{figure} [tb]
\includegraphics[width=6.5cm,height=10cm]{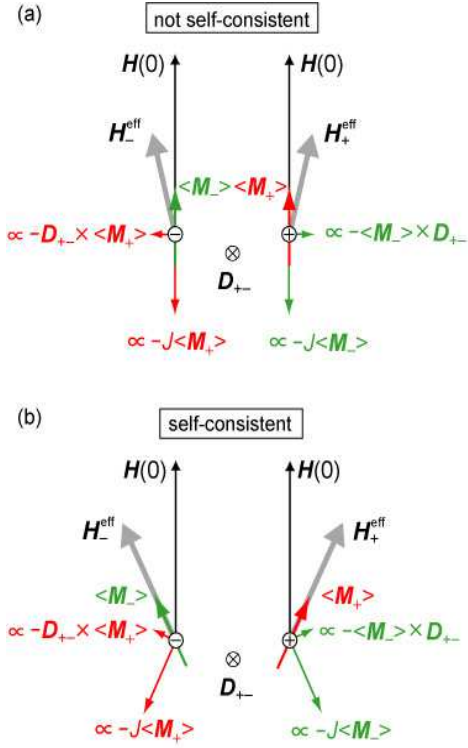}
\caption{\label{Fig9}(Color online) 
	Schematic configuration of the local moments, molecular-fields produced by them, 
and external uniform field for the case of (a) collinear local moments and (b) non-collinear local moments.
	The effective fields {\boldmath $H$}$_+^{\rm eff}$ ({\boldmath $H$}$_-^{\rm eff}$) perceived by the $+$ ($-$)
sublattices are also shown for each case.}

\end{figure}
%%%%%%%%%%%%%%%%%%%Fig9

%%%%%%%%%%%%%%%%%%%%%%%%%%%%%%%%%%%%%%%%%%%%%%%%%%%%%%%%%%%%%%%%%%%%%%%%%%
\subsection{Heisenberg model with DM interaction under magnetic field}

%%%%%%%%%%%%%%%%%%%Fig10
\begin{figure}
\includegraphics[width=8cm,height=11.7cm]{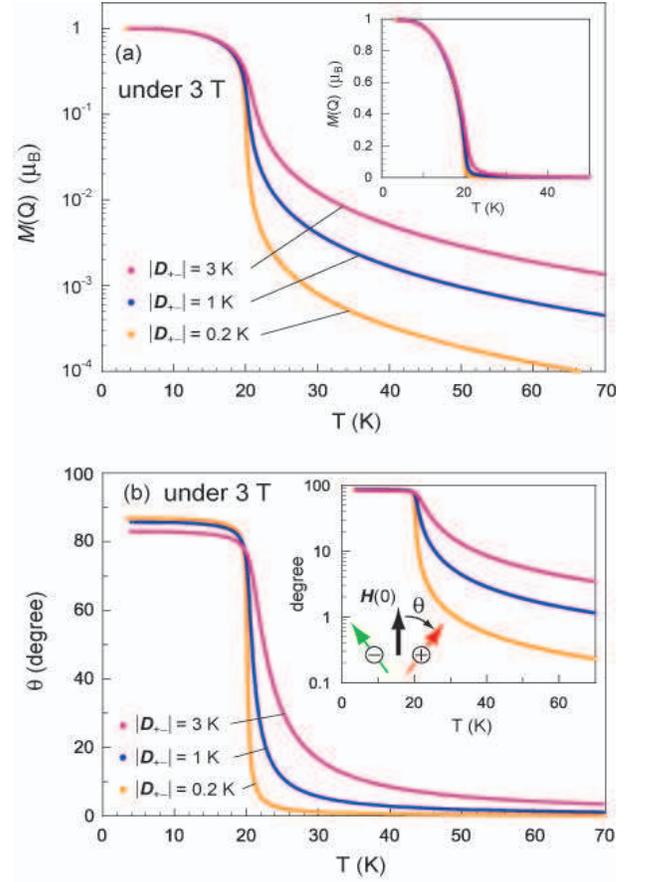}
\caption{\label{Fig10}(Color online) 
	Molecular-field calculations of temperature dependence of (a) the staggered magnetization $M$({\boldmath $Q$}) 
and (b) the non-collinear angle $\theta$. 
	Insets: Plots in (a) linear scale and (b) logarithmic scale.}

\end{figure}
%%%%%%%%%%%%%%%%%%%Fig10

	Next, we consider the Heisenberg model including the DM interaction under a uniform 
field {\boldmath $H$}(0) within the molecular-field approximation. 
	The model was found to exhibit the following characteristic features: (i) a uniform field induces the non-collinear local moments even in the paramagnetic phase (i.e., above the zero-field $T_{\rm N}$); (ii) the divergence of $\chi$({\boldmath $Q$}) 
at $T_{\rm N}$ is suppressed under a uniform field; and (iii) the peak temperature of $\chi$({\boldmath $Q$}) increases with the field in a low-field region. 
	Below we explain these properties in sequence.

	Under a zero field, the self-consistent moments are zero at high temperatures (i.e., paramagnetic), 
while the spontaneous local moments appear below 
the mean-field $T_{\rm N}$ (= $\sqrt{J^{2} + |\Vec{D}_{+-}|^{2}}$). 
	In the paramagnetic state, the local moments are induced under a finite field. 
	Note that such field-induced local moments cannot be collinear, because the collinear state does not satisfy 
$\langle${\boldmath $M$}$_+ \rangle$$\parallel${\boldmath $H$}$_+^{\rm eff}$ and 
$\langle${\boldmath $M$}$_- \rangle$$\parallel${\boldmath $H$}$_-^{\rm eff}$, as illustrated in Fig. \ref{Fig9}(a). 
	Therefore, in a self-consistent result, the field-induced local moments should form a non-collinear 
configuration characterized by positive $M$(0) and positive $M$({\boldmath $Q$})[Fig. \ref{Fig9}(b)] in the whole temperatures. 
	We actually calculated the temperature dependence of the field-induced local moments. 
	Figures \ref{Fig10}(a) and \ref{Fig10}(b) show the behavior of $M$({\boldmath $Q$}) and $\theta$, respectively, under 
$|${\boldmath $H$}(0)$|$ = 3 T with $|${\boldmath $D$}$_{+-}$$|$ = 0.2, 1, and 3 K. 
	Although $M$({\boldmath $Q$}) and $\theta$ show rapid increase around 20 K on cooling, they are finite in the whole temperature range as expected, demonstrating that the AF transition accompanied by thermodynamic singularity does not occur 
in the presence of DM and Zeeman interactions. 
	We also note that as $|${\boldmath $D$}$_{+-}$$|$ increases, the non-collinear feature above $\sim$ 20 K 
becomes more pronounced over a wide temperature range 
[Fig. \ref{Fig10}(b)]. \cite{Ref22}

\begin{figure} [tb]
\includegraphics[width=7.7cm,height=11.7cm]{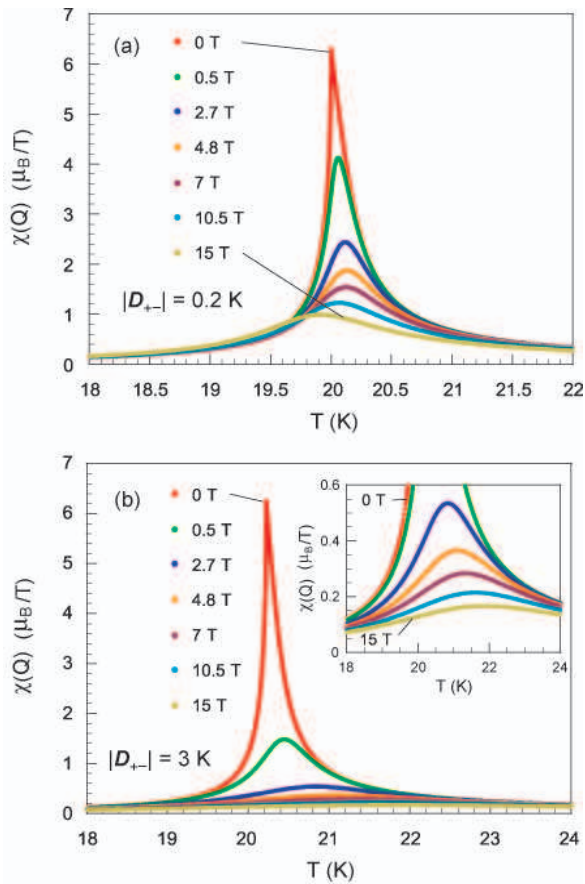}
\caption{\label{Fig11}(Color online) 
	Molecular-field calculations of temperature dependence of $\chi$({\boldmath $Q$}) under various magnetic-fields at   (a) $|${\boldmath $D$}$_{+-}$$|$ = 0.2 and (b) $|${\boldmath $D$}$_{+-}$$|$ = 3 K.
	Inset of (b): Enlarged view of low-$\chi$({\boldmath $Q$}) part.}

\end{figure}

	Figures \ref{Fig11}(a) and \ref{Fig11}(b) show the temperature dependence of $\chi$({\boldmath $Q$}) 
for $|${\boldmath $D$}$_{+-}$$|$ $=$ 0.2 and 3 K, respectively, under various {\boldmath $H$}(0). 
	As {\boldmath $H$}(0) increases, the $\chi$({\boldmath $Q$})-divergence emergent at 0 T is gradually suppressed 
into a rounded peak, consistent with the absence of AF transition. 
	The peak suppression is more prominent for larger $|${\boldmath $D$}$_{+-}$$|$, as seen  
in Fig. \ref{Fig11}(b). 
	These results are in sharp contrast to the magnetic-field effect on $\chi$({\boldmath $Q$}) 
without the DM interaction [Fig. \ref{Fig8}(b)].

%%%%%%%%%%%%%%%%%%%Fig11
%%%%%%%%%%%%%%%%%%%Fig11

%%%%%%%%%%%%%%%%%%%Fig12
\begin{figure} [tb]
\includegraphics[width=6cm,height=12.5cm]{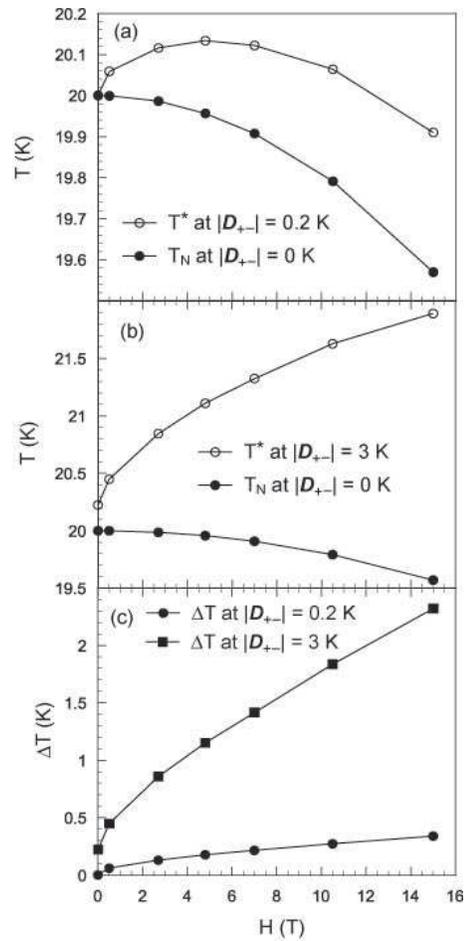}
\caption{\label{Fig12}
	Molecular-field calculations of magnetic-field dependence of $T^{*}$ at (a) $|${\boldmath $D$}$_{+-}$$|$ = 0.2 and
(b) $|${\boldmath $D$}$_{+-}$$|$ = 3 K.
	(c) Molecular-field calculations of magnetic-field dependence of $\triangle$$T$ 
($\equiv$ $T^{*}$ $-$ $T_{\rm N}$) at $|${\boldmath $D$}$_{+-}$$|$ = 0.2 and 3 K.
For comparison, the magnetic-field dependence of $T_{\rm N}$ at $|${\boldmath $D$}$_{+-}$$|$ = 0 K	is also shown 
in (a) and (b).}

\end{figure}
%%%%%%%%%%%%%%%%%%%Fig12

	From the experimental point of view, the peak structure of $\chi$({\boldmath $Q$}) 
[or $\chi$({\boldmath $Q$})-related quantities] has often been interpreted as the indication of AF transition. 
	However, we emphasize again that in the presence of DM interaction the peak under a magnetic field 
should be regarded as the manifestation of the crossover. 
	Therefore we name the peak temperature the crossover temperature, $T^{*}$. 
	The magnetic-field dependence of $T_{\rm N}$ (at $|${\boldmath $D$}$_{+-}$$|$ = 0) and 
$T^{*}$ (at $|${\boldmath $D$}$_{+-}$$|$ $\neq$ 0) is qualitatively different, as shown 
in Figs. \ref{Fig12}(a) and \ref{Fig12}(b): $T^{*}$ increases with {\boldmath $H$}(0) in a low-field region regardless of the strength 
of {\boldmath $D$}$_{+-}$, while $T_{\rm N}$ decreases monotonously. 
	The field-dependence of $T^{*}$ is determined by the superposition of the following two effects. 
	One is the competition between AF exchange and Zeeman interactions, which leads to the suppression of 
$T_{\rm N}$, as is well known.
	The other is the cooperation between the DM and Zeeman interactions, which is naturally expected from the fact that, 
compared with a collinear-AF state, a canted-AF state is favored by both interactions. 
	This cooperation effect appears in $\triangle$$T$ $\equiv$ $T^{*}$ $-$ $T_{\rm N}$, which is plotted against 
$|${\boldmath $H$}(0)$|$ in Fig. \ref{Fig12}(c). 
	In contrast to the case of $T_{\rm N}$, $\triangle$$T$ increases monotonously with $|${\boldmath $H$}(0)$|$,
and this increase is enhanced for larger $|${\boldmath $D$}$_{+-}$$|$ value. 
	The field dependence of $T^{*}$ results from the sum of the two effects and thus may show
non-monotonic field dependence such as shown in Fig. \ref{Fig12}(a).
%

%*****************************************************************************************************
\section{DISCUSSION}
\subsection{Comparison between the experimental and numerical results}

	Now we compare the experimental results with molecular-field calculations. 
	The puzzling magnetism found in $\kappa$-Cl is summarized as follows: (i) even above the zero-field $T_{\rm N}$, 
the field-induced local moments have a staggered component perpendicular to the applied field, and thus the AF transition is not well defined under a magnetic field; (ii) with increasing magnetic field, 
the peak structure in the temperature dependence of 1/$T_{1}$ is suppressed;\cite{Ref17} and (iii) the peak 
temperature in 1/$T_{1}$ increases with the applied field.\cite{Ref17} 
	As seen in Sec. V, the experimental result (i) was reproduced qualitatively by the present calculation. 
	In addition, the calculations showed that the peak structure of 
$\chi$({\boldmath $Q$}) is suppressed with increasing $|${\boldmath $H$}(0)$|$ (Fig. \ref{Fig11}) and that the peak temperature 
increases in a low-field region [Figs. \ref{Fig12}(a) and \ref{Fig12}(b)]. 
	Thus, as far as 1/$T_{1}T$ around $T_{\rm N}$ (or $T^{*}$) reflects well $\chi$({\boldmath $Q$}),\cite{Ref23}  
the molecular-field results also explain features (ii) and (iii) qualitatively. 
	These agreements between the experimental and molecular-field results indicate that the 
puzzling magnetism in $\kappa$-Cl is essentially due to the interplay between Zeeman and DM interactions.
	It should be noted that the NMR line shift in the spin system including the DM interaction is not proportional to the macroscopic magnetization even in the nominally paramagnetic state except the high-temperature regime; 
as was demonstrated by our experimental and numerical results, a magnetic field induces the non-collinear 
spin configuration and thus the NMR line shift reflects not only the uniform component but also
the staggered component of local moments.

%%%%%%%%%%%%%%%%%%%%%%%%%%%%%%%%%%%%%%%%%%%%%%%%%%%%%%%%%%%%%%%%%%%%%%%%%%
\subsection{Quantitative disagreement between the experimental and molecular-field results}

	The agreement between the unconventional magnetism of $\kappa$-Cl and the molecular-field 
results is qualitative, but not quantitative; the features revealed by the calculations are highly enhanced in the experimental observations.
	In $\kappa$-Cl, $D/J$ is considered to be 10$^{-3}$-10$^{-2}$ (Ref. 19); therefore the molecular-field results for $|${\boldmath $D$}$_{+-}$$|$ = 0.2 K (hence $D/J$ = 10$^{-2}$) should be a relevant reference. 
	However, the unconventional features in the numerical results for $|${\boldmath $D$}$_{+-}$$|$ = 0.2 K are not as pronounced as those observed in $\kappa$-Cl.
	First, the values of $\theta$ above $T_{\rm N}$ in the numerical results [Fig. \ref{Fig10}(b)] are too small to be observed experimentally, while the experimental $\theta$ values remain observable even 
well above $T_{\rm N}$ [Fig. \ref{Fig6}(b)].
	Second, the increase in $T^{*}$ with a uniform field is also tiny (0.7 $\%$ at most) in the
calculations [Fig. \ref{Fig12}(a)], while an increase by more than 20 \% was observed in $\kappa$-Cl. \cite{Ref17}
	When these quantitative deviations are considered from the molecular-field point of view,
$D/J$ of $\kappa$-Cl seems as if it were much larger than 10$^{-2}$.
	These quantitative disagreements imply that some sort of effects beyond the molecular-field 
treatment enhances the unconventional features induced by the interplay between the DM and Zeeman 
interactions. 
	Although we have no numerical evidence, we speculate that the most possible candidate is the AF 
short-range order enhanced by low dimensionality, considering that $\kappa$-Cl is actually a quasi-2D
system.
	In fact, the values of $T_{\rm N}/J$ differ significantly for the molecular-field and experimental results: 
$T_{\rm N}/J$ = 1 in the present molecular-field results, while $T_{\rm N}/J$ is less than 0.1 in 
$\kappa$-Cl (the zero-field $T_{\rm N}$ is $\sim$ 23 K and $J$ is considered to be of the order of room temperature). \cite{Ref18} 
	In low-dimensional AF spin systems, the AF short-range order and thus $\chi(\Vec{q}_{\rm AF}$) 
generally grow toward $T_{\rm N}$.
	This growth may enhance the effective magnitude of the DM interaction, resulting in a more prominent 
staggered moment and in more field-dependent $T^{*}$ than expected in the molecular-field calculation.

	The molecular-field results do not reproduce the monotonic decrease in $\chi_{\rm spin}$ down to 
$\sim$ 30 K in $\kappa$-Cl [the inset of Fig. \ref{Fig3}] even in a qualitative level. 
	 This discrepancy is possibly due to the low dimensionality of $\kappa$-Cl, and the further investigation of Eq. (9) is needed beyond the molecular-field approximation. 
	On the other hand, it is likely that the magnetism of $\kappa$-Cl can not be described adequately 
by the model Hamiltonian of Eq. (9), because $\kappa$-Cl is located near the Mott 
transition;\cite{Ref4.1,Ref5,Ref6,Ref7,Ref8,Ref9,Ref10} thus the charge fluctuations are probably not negligible above the critical endpoint at $\sim$ 40 K. 
	Therefore the localized spin model may not be adequate to describe the magnetism of $\kappa$-Cl at high temperatures.
	The Hubbard model, which incorporates the charge degrees of freedom, or the Heisenberg model including
higher-order exchange interactions are considered more realistic.

%%%%%%%%%%%%%%%%%%%%%%%%%%%%%%%%%%%%%%%%%%%%%%%%%%%%%%%%%%%%%%%%%%%%%%%%%%
\subsection{Comparison with the previous studies}

	Our explanation of the field-dependent 1/$T_{1}$ in $\kappa$-Cl is different from the one proposed 
by Hamad \textit{et al}.\cite{Ref17}
	In general, 1/$T_{1}T$ is given by 
\begin{eqnarray}
	 \frac{1}{T_{1}T} \propto \sum_{\Vec {q}}^{} \frac{{\rm Im} \chi(\Vec{q}, \omega_{0})}{\omega_{0}}, 
\end{eqnarray}
where $\omega_{0}$ is an observing NMR frequency proportional to $|${\boldmath $H$}(0)$|$. 
	In most cases, Im $\chi(\Vec{q}, \omega)/\omega$ is $\omega$-independent at low frequencies 
of NMR and hence $1/T_{1}$ does not depend on a magnetic field.
	In order to explain the strongly field-dependent $1/T_{1}$ of $\kappa$-Cl, Hamad \textit{et al}. argued 
that Im $\chi(\Vec{q}, \omega)/\omega$ can have $\omega$-dependence due to slow spin dynamics 
under spin frustration. 
	This is a {\it frequency effect}, but not a {\it field effect} (i.e., the magnetic-field dependence of
$\chi(\Vec{q}, \omega$=0) is not considered). 
	They performed numerical calculations on the Heisenberg model (without the DM interaction) on an anisotropic 
triangular lattice by using the modified spin-wave theory and reproduced the suppression of $1/T_{1}$ peak under a uniform field.
	However their numerical results also exhibit three characteristics inconsistent with the experiments as follows: 
a double-peak structure appears in the temperature dependence of $1/T_{1}$; the peak temperature of $1/T_{1}$ does not depend on 
$|${\boldmath $H$}(0)$|$; and $1/T_{1}$ depends on a magnetic field over a wider temperature range 
than observed. 

		In contrast, we ignore the frequency effect and consider the field dependence of $\chi(\Vec{q}, \omega$=0) in the presence of DM interaction.
	This is the field effect. 
	Our treatment exhibits the strong field dependence of $\chi(\Vec{Q})$ only around the zero-field $T_{\rm N}$
[Figs. \ref{Fig11}(a) and \ref{Fig11}(b)].
	As described in Sec. VI A, as far as $\chi(\Vec{q}_{\rm AF})$ and $1/T_{1}T$ are well correlated near 
$T_{\rm N}$ (or $T^{*}$) in $\kappa$-Cl, \cite{Ref23} the molecular-field results explain the experimental results 
without the discrepancies encountered in the model postulated by Hamad \textit{et al}.\cite{Ref17}
	Thus, the unconventional field dependence of $1/T_{1}$ observed only around the zero-field $T_{\rm N}$ 
is successfully understood in terms of the {\it field effect} on $\chi(\Vec{q}_{\rm AF})$.

	The field-induced staggered moment, which is a key 
consequence of the present study, has often been observed and discussed in quantum disordered or gapped spin systems. 
	Oshikawa and Affleck\cite{Ref24} argued that in S = 1/2 AF spin chains, the interplay between the Zeeman and DM 
interactions induces the staggered field perpendicular to the applied magnetic field, resulting in the field-induced 
staggered moment.  
	This actually explains the observations in Cu benzoate \cite{Ref25} and copper pyrimidine dinitrate.\cite{Ref26} 
	The spin-gapped system with DM interaction, such as NENP ($S$ = 1 AF spin chain) \cite{Ref27,Ref28} 
and SrCu$_{2}$(BO$_{3}$)$_{2}$ ($S$ = 1/2 2D frustrated spin system),\cite{Ref29,Ref30} also show a field-induced staggered moment below a critical field. 
	These properties are also well understood in terms of the staggered-field mechanism. 
	The present study is addressed as the extension of this issue to the quasi-2D spin systems undergoing 
a finite-temperature AF transition. 
	Recently, the NMR line shift of La$_{2}$CuO$_{4}$, in which the DM interaction is inherent, has been discussed 
in terms of the field-induced staggered moment.\cite{Ref31}

%***********************************************************************
\section{CONCLUSION}

	We performed the magnetization and NMR measurements on $\kappa$-(BEDT-TTF)$_{2}$Cu[N(CN)$_{2}$]Cl for 
deeper understanding of its magnetism. 
	We found that the application of a uniform field induces a staggered moment perpendicular to the applied field even above $T_{\rm N}$. 
	The molecular-field calculation revealed that the field-induced staggered moment comes from the interplay 
between the Zeeman and DM interactions and that this interplay explains the several otherwise puzzling magnetic features
observed in experiments. 
	By comparing the experiment results with our numerical results, we concluded that the DM interaction 
affects significantly the field-induced magnetism in a paramagnetic phase especially in low-dimensional systems.

\section*{Acknowledgements}
	The authors thank M. Takigawa, S. Todo, S. Miyahara, and M. Imada for fruitful discussion and S. Niitaka for
his support of SQUID measurements.
	This work was in part supported by a Grants-in-Aid for Scientific Research in Priority Areas of ``Molecular Conductors" 
(Grant No. 15073204) and ``Physics of Superclean Materials" (Grant No. 17071003)
from the Ministry of Education, Culture, Sports, Science and Technology and a Grant-in-Aid for Scientific 
Research (Grants No. 15104006, No. 20244055, and No. 20540346) from the JSPS.


\begin{thebibliography}{99}
\bibitem{Ref21} Strictly speaking, our numerical results show not the genuine divergence of $\chi$({\boldmath $Q$}) 
but a finite peak value of $\sim$ 6.28 $\mu_B$/T, because $\chi$({\boldmath $Q$}) was calculated under a finite 
value of {\boldmath $H$}({\boldmath $Q$}). In our numerical accuracy, the divergence of $\chi$({\boldmath $Q$}) appears as the peak value lager than 6.2 $\mu_B$/T.
\bibitem{Ref22} Concerning the magnetic-field effect on the non-collinear feature, we also found that $\theta$ above the mean-field $T_{\rm N}$ is almost field-independent 
between under 0.1 T and 3 T, while $|\langle${\boldmath $M$}$_+ \rangle|$ is nearly proportional to {\boldmath $H$}(0) in this field range (not shown).
\bibitem{Ref23} Around $T_{\rm N}$ (or $T^{*}$), 1/$T_{1}T$ is generally dominated by 
Im$\chi$({\boldmath $q$}$_{\rm AF}$,$\omega_{0})/\omega_{0}$, where {\boldmath $q$}$_{\rm AF}$ and $\omega_{0}$ represent 
the wave-number vector of AF order and the observing NMR frequency, respectively. 
Within the time-dependent Ginzburg-Landau equation, the sign of 
$\frac{\partial}{\partial T} \chi(\Vec{q}_{\rm AF},0)$ 
and $\frac{\partial}{\partial H} \chi(\Vec{q}_{\rm AF},0)$ agrees with, respectively, that of 
$\frac{\partial}{\partial T}$Im$\chi$({\boldmath $q$}$_{\rm AF}$,$\omega_{0})/\omega_{0}$ 
and $\frac{\partial}{\partial H}$Im$\chi$({\boldmath $q$}$_{\rm AF}$,$\omega_{0})/\omega_{0}$, 
as far as $\omega_{0}$ is lower enough than the characteristic relaxation frequency of the electron-spin dynamics. 
Thus it is likely that around $T_{\rm N}$ (or $T^{*}$), $\chi(\Vec{q}_{\rm AF}$) and 1/$T_{1}T$ show the qualitatively similar temperature dependence and magnetic-field dependence.
\end{thebibliography}
\end{document}